\documentclass[aps,prl,twocolumn,groupedaddress]{revtex4}
\usepackage{graphicx}

\begin{document}
\newcommand{\bea}{\begin{eqnarray}}
\newcommand{\eea}{\end{eqnarray}}
\newcommand{\beq}{\begin{equation}}
\newcommand{\eeq}{\end{equation}}
\newcommand{\lav}{\langle}
\newcommand{\rav}{\rangle}
\def \tr{{\mbox{tr~}}}
\def \ra{{\rightarrow}}
\def \ua{{\uparrow}}
\def \da{{\downarrow}}
\def \be{\begin{equation}}
\def \ee{\end{equation}}
\def \ba{\begin{array}}
\def \ea{\end{array}}
\def \bea{\begin{eqnarray}}
\def \eea{\end{eqnarray}}
\def \nn{\nonumber}
\def \l{\left}
\def \r{\right}
\def \half{{1\over 2}}
\def \etal{{\it {et al}}}
\def \cH{{\cal{H}}}
\def \cM{{\cal{M}}}
\def \cN{{\cal{N}}}
\def \cQ{{\cal Q}}
\def \cI{{\cal I}}
\def \cV{{\cal V}}
\def \cG{{\cal G}}
\def \cF{{\cal F}}
\def \cZ{{\cal Z}}
\def \bS{{\bf S}}
\def \bI{{\bf I}}
\def \bL{{\bf L}}
\def \bG{{\bf G}}
\def \bQ{{\bf Q}}
\def \bK{{\bf K}}
\def \bR{{\bf R}}
\def \br{{\bf r}}
\def \bu{{\bf u}}
\def \bq{{\bf q}}
\def \bk{{\bf k}}
\def \bz{{\bf z}}
\def \bx{{\bf x}}
\def \bpsi{{\bar{\psi}}}
\def \tJ{{\tilde{J}}}
\def \W{{\Omega}}
\def \e{{\epsilon}}
\def \lam{{\lambda}}
\def \L{{\Lambda}}
\def \a{{\alpha}}
\def \t{{\theta}}
\def \b{{\beta}}
\def \g{{\gamma}}
\def \D{{\Delta}}
\def \d{{\delta}}
\def \w{{\omega}}
\def \s{{\sigma}}
\def \f{{\varphi}}
\def \x{{\chi}}
\def \e{{\epsilon}}
\def \h{{\eta}}
\def \G{{\Gamma}}
\def \z{{\zeta}}
\def \hatt{{\hat{\t}}}
\def \hn{{\bar{n}}}
\def \vk{{\bf{k}}}
\def \vq{{\bf{q}}}
\def \gk{{\g_{\vk}}}
\def \nd{{^{\vphantom{\dagger}}}}
\def \yd{^\dagger}
\def \av#1{{\langle#1\rangle}}
\def \ket#1{{\,|\,#1\,\rangle\,}}
\def \bra#1{{\,\langle\,#1\,|\,}}
\def \braket#1#2{{\,\langle\,#1\,|\,#2\,\rangle\,}}

\title{Noise Correlations in one-dimensional systems of ultra-cold fermions}
\author{L. Mathey$^1$, E. Altman$^{2}$ and A. Vishwanath$^{3}$}
\affiliation{$^1$Physics Department, Harvard University, Cambridge, MA 02138 \\
$^2$ Department of Condensed Matter Physics, The Weizmann
Institute of Science Rehovot, 76100, Israel \\
$^3$Department of Physics, University of California, Berkeley, CA
94720}
\date{\today}

\begin{abstract}
Time of flight images reflect the momentum distribution of the atoms
in the trap, but the spatial noise in the image holds information on
more subtle correlations. Using Bosonization, we study such noise
correlations in
generic one dimensional systems of ultra cold
fermions. Specifically, we show how pairing as well as spin and
charge density wave correlations may be identified and extracted
from the time of flight images. These incipient orders manifest
themselves as power law singularities in the noise correlations,
that depend on the Luttinger parameters,
which suggests a general experimental technique to obtain them.
\end{abstract}


\maketitle

Recent advances in trapping and control of ultra cold atoms prompted
the realization of truly one dimensional systems. Important
benchmarks, such as the Tonks-Girardeau gas \cite{paredes,weiss} and
the Mott transition in one dimension\cite{stoeferle}, have been
achieved by trapping bosonic atoms in tight tubes formed by an
optical lattice potential. Novel transport properties of one
dimensional lattice bosons have been studied using these
techniques\cite{fertig}. More recently, a strongly interacting one
dimensional Fermi gas was realized using similar trapping
methods\cite{Moritz}. Interactions between the fermion atoms
were controlled
by tuning a Feshbach resonance in these experiments.

Even the simplest model of interacting spinless fermions in one
dimension exhibits
different regimes of behavior. Although spinless fermions confined
to an isolated line display a single thermodynamic phase, it is
convenient to characterise different parameter regimes in terms of
which susceptibilities are divergent. A divergent susceptibility in
a particular channel implies that coupling independent one
dimensional systems in that channel will lead to an instability. If
we classify 'phases' of the one dimensional system in this way, then
the 'phase' diagram of spinless fermions includes a diverging charge
density wave (CDW) susceptibility for repulsive interaction and
diverging superconducting (SC) susceptibility for attractive
interaction. In the case of spin-1/2 Fermions, which can exist
either in a spin gapped or gapless phase in the continuum, a much
richer competition of 'phases' (as defined from divergent
susceptibilities) occurs. These include both triplet SC (TSC) and
singlet SC (SSC) as well as CDW and spin density wave (SDW) (see
e.g. \cite{review}). Numerous proposals were given for realization
of a variety of different phases in ultra cold Fermi systems
\cite{theory1} as well as Bose-Fermi
mixtures\cite{cazalilla,mathey}. However methods for detection and
experimental characterization of all of these strongly correlated
states are lacking.

There is a fundamental difficulty, not confined to atomic systems,
in applying a unified probe of correlations. A standard detection
scheme in solid state systems involves coherent scattering of
external particles, such as neutrons or photons, that naturally
couple to the system (weakly) via the particle hole channel. Such a
probe is therefore sensitive to spin and charge order, but not to
superconductivity. Thus parts of the phase diagram of one
dimensional systems, which in theory are dual to each other, are not
treated equally by such probes.
In this paper we show that by measuring the noise correlations in time of flight images
one can gain access to, and distinguish between different
power-law particle-particle correlations, such as TSC and SSC, and
particle-hole correlations such as CDW and SDW.

Atomic shot noise in time of flight imaging was proposed  in Ref.
\cite{ehud} as a probe of many body correlations in systems of ultra
cold atoms.
Specifically, what may be
measured using this approach is the momentum space correlation
function of the atoms in the trap
\be G_{\a\a'}(k,k')=\av{n_{\a k}
n_{\a'k'}}-\av{n_{\a k}}\av{n_{\a' k'}}, \label{G}
\ee
where $\a,\a'$ are spin indices.
The method was successfully applied to experiments measuring correlations in
both particle-particle and particle hole channels.
For example Greiner \etal \cite{markus} measured pairing correlations in an
expanding cloud of atoms originating from dissociated molecules.
Other experiments measured density correlations in insulating
states of lattice bosons \cite{BlochHBT,Spielman} and fermions\cite{Rom}.
A related effect
corresponding to the classical limit of the noise correlations was observed
in Ref. ~\cite{zoran}.

The analysis
In the examples considered in \cite{ehud}, calculation of the noise
correlations (\ref{G}) was greatly simplified by the assumption of a
mean field BCS state and a perfect Mott insulator respectively.
However, mean field theory fails to describe the one dimensional
systems discussed here. Instead the long distance behavior of
correlations  is described by the Luttinger liquid framework, which
yields power law decays, determined by one or two Luttinger
parameters. We shall calculate the correlation function (\ref{G})
within this framework, and extract its leading singularities. These
turn out to be directly related to the pure CDW, SDW and SC
correlations.

{\em Spinless fermions --} Let us start by treating in detail the
simplest case, namely that of spinless fermions in the continuum.
Later we shall generalize to spin-$\half$ fermions. To implement a
Luttinger liquid of spinless fermions one can use fully polarized
fermionic atoms. The interaction may be tuned via Feshbach resonance
in an odd angular momentum channel (e.g. $p$-wave). Another
possibility for tuning interactions is to use a Bose-Fermi mixture,
where phonons in the bose superfluid mediate effective interactions
between the fermions\cite{mathey}.

\begin{figure}[t]
\includegraphics[width=8cm]{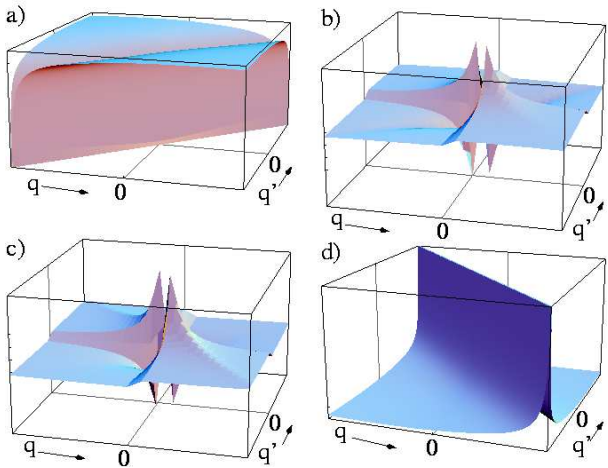}
\caption{ The noise correlation function (\ref{CorrInt}) of a
spinless Luttinger liquid for values of the Luttinger parameter $K$
a) -- d): $K=0.4$, $0.8$, $1.25$ and $2.5$. For $K<1/2$ there are
strong anti correlations (reflecting the particle-hole nature of
CDW), which appear as a power law singularity on the line $q-q'=0$.
For $K>2$ correlations switch to particle-particle (SC) and appear
as a powerlaw singularity on the line $q+q'=0$. For $1/2<K<2$ we
find a singularity only when the two momenta approach the two fermi
points (i.e $q=q'=0$). The system shows a signature of both
particle-hole and particle-particle correlations with the sign of
correlation depending on the directions of approach to the fermi
points.} \label{fig:f-nospin}
\end{figure}

Regardless of the microscopic interactions, the low energy physics
of one dimensional spinless fermions is described by a universal
harmonic theory, the Luttinger liquid. The action, describing
phonons in the Fermi system, is written in terms of the phase field
$\Phi$ or its dual $\Theta$: \be S=\frac{K}{2\pi} \int d^2
x\partial_\mu\Phi\partial^\mu\Phi = \frac{K^{-1}}{2\pi} \int d^2 x
\partial_\mu\Theta\partial^\mu\Theta.
\label{S0} \ee Here $\mu=0,1$, $x_0=v_F\tau$ and $K$ is the
Luttinger parameter. $K>1$ ($K<1$) corresponds to attractive
(repulsive) interactions. To calculate physical correlation
functions one needs the bosonization identity connecting the
oscillator fields to the physical fermions\cite{haldane}: \be
\psi_{\pm}(x)=\frac{1}{\sqrt{2\pi\a}}e^{\pm i\Theta(x)+i\Phi(x)},
\label{bos} \ee where  $\a$ is an infinitesimal to be taken to zero
at the end of any calculation.
 $\psi_\pm(x)$ correspond to the slowly varying, left and right
moving components of the fermion field $\psi(x)=e^{-ik_f
x}\psi_-(x)+e^{ik_f x}\psi_+(x)$. The commutation relation of the
oscillator fields is given by: \bea [\Theta(x), \Phi(0)] & = &
\frac{1}{2} \log\frac{\alpha + i x}{\alpha - i x}. \label{com} \eea
The relevant order parameters can be expressed simply in terms of
the slow fields. The $2k_f$ charge density wave operator order
parameter is given by $O_{CDW}(x)=\e^{2ik_fx}\psi\yd_+\psi\nd_-$,
while the SC order parameter is $O_{SC}(x)=\psi_+\psi_-$. A standard
calculation\cite{review} using (\ref{S0}) and (\ref{bos}) yields
$\av{O\yd_{CDW}(x)O\nd_{CDW}(0)}\sim \cos(2k_f)|x|^{-2K}$ and
$\av{O\yd_{SC}(x)O\nd_{SC}(0)}\sim |x|^{-2/K}$. Taking the spatial
and temporal Fourier transform one can see that CDW and SC
susceptibilities diverge when $K<1$ and $K>1$ respectively.

We are interested in the equal time correlation function (\ref{G})
with $k$ and $k'$ close to the opposite fermi points at $\pm k_f$.
The operator $\av{n_k n_{k'}}$, with $k$ ($k'$) near the fermi
points $k_f$ ($-k_f$) is given by: \bea
\av{n_k n_{k'}} &=&\frac{1}{V}\int \prod_{i=1}^4 dx_i e^{iq (x_1-x_2)+i q' (x_3-x_4)}\nn\\
&&\av{\psi\yd_+(x_1) \psi\nd_+(x_2)\psi\yd_-(x_3) \psi\nd_-(x_4)},
\label{nknk'} \eea where $q=k-k_f$ and $q'=k'+k_f$ denote the
momenta relative to the respective fermi points. Using the
bosonization identity (\ref{bos}) and the commutation relation
(\ref{com}) with the harmonic action (\ref{S0}) we obtain for the
noise correlation function (\ref{G}):

\bea G(q, q') & = & \int\frac{e^{iqx_{12}+iq'x_{34}}}{(2\pi)^2 V}
\mathcal{F}(x_{12})\mathcal{F}^\star(x_{34})(\mathcal{A} -1)
\label{CorrInt}. \eea Where $x_{ij}=x_i-x_j$ the integration is over
$x_{12}, x_{34}, x_{23}$, and \bea \mathcal{F}(x) & \equiv &
\Big(\frac{a^2}{a^2+x^2}\Big)^g
\frac{1}{\alpha - i x}\nn\\
\mathcal{A} & \equiv & \left[\frac{(a^2+x_{14}^2)(a^2+x_{23}^2)}
{(a^2 + x_{13}^2)(a^2 + x_{24}^2)}\right]^h \label{F} \eea with $a$
a short distance cutoff (which will be kept finite in contrast to
 $\alpha$,
 see e.g. \cite{review}), and exponents $g=(K+K^{-1}-2)/4$ and
$h=(K-K^{-1})/4$. We shall undertake a full numerical integration of
(\ref{CorrInt}). But first let us obtain the leading singularities
in the limits $K\ll 1$, $K\gg 1$, and
 $|1-K|\ll 1$. We note that (\ref{nknk'}) 'contains'
both the CDW and SC correlations. For example, the contribution to
(\ref{CorrInt}) from setting $x_1=x_4\equiv x$ and $x_2=x_3\equiv
x'$ is simply the fourier transform of the CDW correlation at
$q-q'$, with a minus sign due to anticommutation of the Fermi
fields. For $K\ll 1$ since CDW correlations decay very slowly, we
expect that there are no singular contributions arising from the
deviations $x_{14}$, $x_{23}$, so that the CDW correlations dominate
the integral. Indeed, $x_{14}$ and $x_{23}$ appear  in the numerator
of $\mathcal{A}$, and since for $K\ll 1$, $h$ is very negative, the
contribution of large deviations is effectively suppressed. This
gives $G(q,q')\propto -|q-q'|^{2K-1}$. In a similar manner setting
$x_1=x_3$ and $x_2=x_4$ yields the SC correlation at $q+q'$. For
$K\gg 1$, $h$ is large and positive, effectively suppressing the
deviations $x_{13}$ and $x_{24}$. Thus in this limit we find
$G(q,q')\propto |q+q'|^{2/K-1}$. With these considerations we would
not expect a singularity in $G(k,k')$ in the region $\half<K<2$ due
to SC or CDW correlations. However, the approximation presented
above is not valid for such moderate values of $K$. Within this
regime we can obtain analytic results for the leading singularities
of (\ref{CorrInt}) in the limit $|K-1| \ll 1$ ($|h| \ll 1$).  The
idea is to expand $G(q,q')$ to powers of $h$. The term linear in $h$
vanishes identically, and to quadratic order: \be G(q,q')\sim
-h^2{\rm sgn}(q){\rm sgn}(q'){\rm min}\left({1\over |q|},{1\over
|q'|} \right) \label{Gh2} \ee

We now turn to integrate (\ref{CorrInt}) numerically in order to
obtain $G(q,q')$. This is carried out on a finite size system with
periodic boundary conditions so that $x_{ij}$ in (\ref{CorrInt}) are
replaced by $(L/2\pi)\sin(2\pi x_{ij}/L)$. The results, summarized
in Fig. \ref{fig:f-nospin}, indicate that the anticipated power law
singularities persist in the entire region $K>2$ and $K<1/2$, where
the equal time order parameter correlations yield a singular
contribution. The structure of the correlations in the regime
$1/2<K<2$ is consistent with the analysis presented above for
$K\approx 1$.

\begin{figure}[t]
\includegraphics[width=5cm]{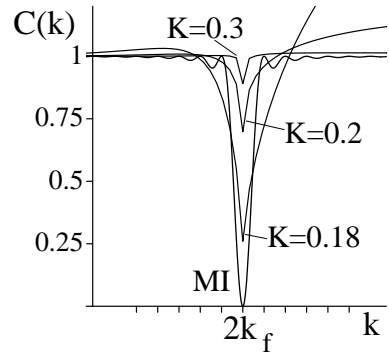}
\caption{
 The normalized correlation function (\ref{Ck}) for a realistic finite system
 with $N\approx 100$ particles per 1d tube, as a function
 of momentum $k$ in the vicinity of $2 k_f$. (The
 scale of the
 horizontal axis is indicated by the marks which correspond to increments
 of $2\pi/L$.)
  The result is plotted for $K=0.3,0.2,0.18$ in the CDW regime.
 The broad dip due to the power-law singularity should be contrasted with the
 sharper dip for the case of a finite Mott insulating state (MI).
} \label{fig:Ck}
\end{figure}

To predict the visibility of the peaks we consider the normalized
correlation function typically used in such
experiments\cite{BlochHBT,Rom} \bea \label{Ck} C(k) & = & \frac{\int
dp \av{n_{p-k/2} n_{p+k/2}}}{\int dp \av{n_{p-k/2}}\av{n_{p +k/2}}}
\eea The numerator is given by the connected part of the correlation
function (\ref{CorrInt}), i.e. the part containing $\mathcal{A}$,
while the denominator is the disconnected part. Fig. \ref{fig:Ck}
shows a computation of $C(k)$ for a realistic system of decoupled
tubes with $100$ particles per tube. Thus the short range cut-off,
which appears in Eqs. (\ref{F}) is set to $a=2\pi/k_f=L/100$. The
momenta are then given in units of $2\pi/L$, which appears as the
scale on the horizontal axis. The presence of QLRO, rather than LRO,
is evident by the appearance of broad power-law dip in $C(\bk)$
compared to the sharp dip in the case of perfect order. We note that
the distinct dip is clearly observable only for strong interactions
(i.e. $K\lesssim0.25$).

{\em Spin-$\half$ fermions --} A spin-$\half$ system exhibits much
richer physics than the spinless one, and it may be realized using
interactions in the $s$-wave channel. The low energy physics is
captured by the universal action: \bea S &=& \sum_\nu\frac{1}{2 \pi
K_\nu}\int  dx d\tau\left[\frac{1}{v_\nu}(\partial_\tau\theta_\nu)^2
+
v_\nu (\partial_x\theta_\nu)^2\right]\nn\\
&&+\frac{2 g_{1\perp}}{(2\pi\a)^2}\int dx d\tau
\cos(\sqrt{8K_\s}\theta_\s) \label{SLL} \eea Where $\nu=\rho,\sigma$
correspond to the charge and spin degrees of freedom, and
$\Theta_{\rho,\s}=(\Theta_\ua\pm\Theta_\da)/\sqrt{2}$ and
$\Phi_{\rho,\s}=(\Phi_\ua\pm\Phi_\da)/\sqrt{2}$. The nonlinear term
that arises in the spin channel is due to back scattering. The
bosonization formula is identical to (\ref{bos}) with Fermi as well
as oscillator fields taking a spin index.

The low energy fluctuations of the spin-$\half$ LL are dominated by
four order parameters constructed from the slow fermion fields:
$O_{CDW}=\sum_s e^{2ik_f x}\psi\yd_{-,s}\psi\nd_{+,s}$,
$O_{SDW}=\sum_s\e^{2ik_fx}\psi\yd_{-,s}\psi\nd_{+,-s}$
$O_{SSC}=\sum_{ss'}\e_{ss'}\psi\nd_{-,s}\psi\nd_{+,s'}$, and
$O_{TSC}=\psi\nd_{-,\ua}\psi\nd_{+,\ua}$ (we chose the $S_z=1$
component for convenience, the other components must be degenerate
by $su(2)$ spin symmetry).
Calculation of the correlation functions associated with these
operators is more subtle than in the spinless case due to the
presence of the nonlinear term in (\ref{SLL}). It was carried out in
Ref. \cite{giamarchi} using RG analysis.

The
 resulting
phase diagram, in the sense of diverging susceptibilities, is
 of the following structure:
 For positive backscattering $g_{1\perp}>0$ flows to $0^+$,
$K_\s\to 1$, and $K_\rho$ does not flow. To derive the power law
decay of order parameter correlations one can assume the fixed-point
values of the Luttinger parameters. The result is $1/x^{K_\rho+1}$
for CDW/SDW and $1/x^{K_\rho^{-1}+1}$ for SSC/TSC. This implies
equally diverging susceptibilities of CDW and SDW for $K_\rho<1$ and
of SSC and TSC for $K_\rho>1$. The degeneracy between order
parameters is removed by logarithmic corrections favoring SDW and
TSC\cite{giamarchi}. In the opposite regime $g_{1\perp}<0$, the non
linear term is relevant, opening a spin gap in the spectrum.
Consequently SDW and TSC correlations are short ranged. The scaling
exponents of the remaining CDW and SSC correlations may be found by
formally setting $K_\s\to 0$ \cite{review}, giving $R(x)\sim
|x|^{-K^{\pm 1}}$ for CDW/SSC respectively. Hence there is a
sizeable region $\half<K_\rho<2$, where both susceptibilities are
divergent.

\begin{figure}[t]
\includegraphics[width=8.5cm]{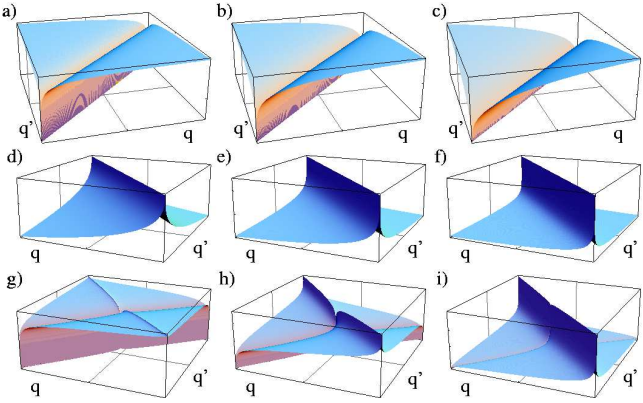}
\caption{ Noise correlations of a spin-$1/2$ fermion Luttinger
liquid
 with $g_{1\perp}<0$. Different rows correspond
to different spin dependent correlations: (a,b,c) $G_{\ua\ua}$,
which in the spin gapped phase is sensitive only to CDW
correlations; (d,e,f) $G_{\ua\da}$, sensitive only to the SSC
correlations; (g,h,i) $G=\sum_{\a\b}G_{\a\b}$ sensitive to both CDW
and SSC. Columns correspond to different values of the Luttinger
parameter $K_\rho$ ($K_\s$ is set to zero in the spin gapped phase):
(a, d, g) $K_\rho=0.8$; (b, e, h) $K_\rho=1$; (c, f, i)
$K_\rho=1.25$. The noise correlations in $G$ clearly show
coexistence of the two (quasi) orders  for $1/2<K<2$ (g,h,i).}
\label{fig:spingap}
\end{figure}

The expressions for the spin-resolved
 noise correlation functions $G_{\uparrow\uparrow}(q,q')$ and $G_{\ua\da}(q,q')$
 are  identical to Eq. (\ref{CorrInt}), with the exponents in
Eq. (\ref{F}) given by: \bea g & = & (K_\rho+K_\rho^{-1})/8+
(K_\sigma+K_\sigma^{-1})/8-1/2\nn\\
h_{\uparrow\uparrow} & = & (K_\rho-K_\rho^{-1})/8+
(K_\sigma-K_\sigma^{-1})/8\label{hupup}\nn\\
h_{\ua\da} & = & (K_\rho-K_\rho^{-1})/8-
(K_\sigma-K_\sigma^{-1})/8\label{hdownup} \label{gh} \eea

As before, we note that this complicated function contains, along
with other contributions, the equal time correlations of the four
order parameters discussed above. For example $G_{\ua\ua}$ contains
the TSC and a component of the CDW correlations, whereas
$G_{\ua\da}$ contains the SDW, as well as components of both the
singlet and triplet SC. Thus, $G=\sum_{ss'}G_{ss'}$, contains all
the order parameter correlations equally. As in the spinless case we
expect that when these correlations decay slowly enough they give
the most singular contribution to $G_{ss'}$. Where they are not
singular, other contributions may take the lead.

Let us first consider the case $g_{1\perp}>0$ taking the values of
the parameters at the fixed point: $K_\s\to 1$ and $g_{1\perp}\to
0$. We note, that in contrast with the spinless case, the equal time
order parameter correlations decay too fast to give a singular
contribution to $G_{ss'}$. Exactly as in the spinless case we can
obtain analytic expressions for the leading contributions to
$G_{ss'}$ for $K_\rho\approx 1$. Expanding to quadratic order in
$h_{ss'}$ we again find $G_{ss'}(q,q')$ given by Eq. (\ref{Gh2})
with $h_{ss'}$ replacing $h$. This general behavior agrees with the
result of numerical integration of $G_{ss'}(q,q')$, although there
may be corrections to the power of the singularities away from the
line $K_\rho=1$.

In the opposite regime of negative backscattering one should
evaluate the integral (\ref{CorrInt}) with the exponents given by
(\ref{gh}) in the limit $K_\s\to 0$. Taking, for example,
$G_{\ua\ua}$ we see that $h_{\ua\ua}\to-\infty$ effectively
enforcing $x_{14}\pm x_{12}=0$. The remaining integral over $x\equiv
(x_{12}-x_{34})/2$ then depends only on the combination
$4(g+h_{\ua\ua})+2=K_\rho$. We find $G_{\ua\ua}\sim -{\rm
sgn}(1-K_\rho) |q-q'|^{K_\rho-1}$ and $G_{\ua\da}\sim {\rm
sgn}(1-1/K_\rho) |q-q'|^{1/K_\rho-1}$. These are precisely the
expected contributions from equal time CDW and SSC order parameter
correlations. In the region $1/2<K_\rho<2$ where the
susceptibilities of both order parameters diverge, we find a
divergence in the noise correlations due to only one of the orders
and a sharp cusp due to the other incipient order.
 Specifically, for $1/2<K_\rho<1$, we find
 a divergence in $G_{\ua\ua}$, and a cusp in $G_{\ua\da}$,
 whereas for $1<K_\rho<2$,
 we find a cusp in $G_{\ua\ua}$, and a divergence in $G_{\ua\da}$.


We also carry out a numerical integration of (\ref{CorrInt}). The
results are displayed in Fig. (\ref{fig:spingap}). They are in
perfect agreement with the analytic forms discussed above. Presence
of a spin gap is immediately seen by inspecting the different
components of $G_{ss'}$. Specifically, particle-particle
correlations arise {\em only} in the $\ua\da$ channels, therefore
they correspond to $SSC$. Similarly particle-hole correlations are
seen in the $\ua\ua$ and not in the $\ua\da$ channel, consistent
with CDW but not SDW quasi long range order.

In summary, we propose that analysis of the spatial noise
correlations in time of flight images, may serve as a detailed probe
of correlations in one dimensional systems of ultra-cold
fermions\cite{Moritz}. In contrast with Bragg scattering, which can
only detect correlations in the particle-hole channel (CDW/SDW),
such a probe would be equally sensitive to superfluid correlations
in the particle-particle channel. The power-law singularities in the
noise correlations are directly related to the singularities of the
equal time order parameter correlations, hence they allow to extract
Luttinger parameters from experiments. In addition we find new
singularities in the noise correlation function, pointing to the
ground state structure, even where the pure equal time order
parameter correlations are non singular.

{\em Acknowledgements.} Fruitful discussions with E. Demler, M. D.
Lukin, and T. Esslinger are gratefully acknowledged. This research
was supported by US-Israel Binational Science Foundation (E. A.), A.
P. Sloan foundation and DOE LDRD DEAC0376SF00098 (A. V.).

\end{document}